\documentclass[sigconf,screen]{acmart}

\usepackage{booktabs}
\usepackage{listings}
\usepackage{multirow}
\usepackage{microtype}
\usepackage{url}
\usepackage{subcaption}
\usepackage{balance}
\usepackage{tikz}
\usetikzlibrary{arrows.meta}

\lstdefinestyle{cfix}{
  language=C,
  basicstyle=\scriptsize\ttfamily,
  basewidth={0.5em,0.45em},
  commentstyle=\color[HTML]{5F5E5A},
  keywordstyle=\bfseries,
  showstringspaces=false,
  columns=fixed,
  keepspaces=true,
  breaklines=true,
  xleftmargin=0pt,
  aboveskip=4pt,
  belowskip=4pt
}

\setcopyright{cc}
\setcctype{by}
\acmDOI{10.1145/3842651.3843185}
\acmYear{2026}
\copyrightyear{2026}
\acmISBN{979-8-4007-2969-0/2026/10}
\acmConference[FTA '26]{Proceedings of the 1st International Workshop on Firmware Testing and Analysis}{October 4--9, 2026}{Oakland, CA, USA}
\acmBooktitle{Proceedings of the 1st International Workshop on Firmware Testing and Analysis (FTA '26), October 4--9, 2026, Oakland, CA, USA}
\acmSubmissionID{splashws26ftamain-p56-p}
\received{2026-07-21}
\received[accepted]{2026-08-09}

\begin{document}

\title{From Silicon to Boot Code: Extending Automated Program Repair to Firmware-Layer Security Workarounds}

\author{Maisha Mastora}
\correspondingauthor
\orcid{0009-0003-4799-5658}
\affiliation{%
  \institution{University of New Hampshire}
  \city{Durham}
  \country{USA}
}
\email{maisha.mastora@unh.edu}

\author{Dean Sullivan}
\orcid{0000-0002-7186-4346}
\affiliation{%
  \institution{University of New Hampshire}
  \city{Durham}
  \country{USA}
}
\email{dean.sullivan@unh.edu}

\begin{abstract}
Automated program repair (APR) research has been constrained to design time. Current techniques localize and fix bugs in Register-Transfer Level (RTL) or High Level Synthesis (HLS) designs before a chip reaches production. Once a hardware vulnerability surfaces post-silicon, the patch content must be manually generated. Existing automation methods address patch deployment
but not patch synthesis. Based on established pre-silicon APR methods, we study the feasibility of extending to this firmware layer a dictionary-guided, localize-synthesize-validate APR methodology originally developed for RTL
repair. An automated commit-clustering miner surfaces recurring fix templates
across the EDK~II (UEFI) firmware repository's full commit history without depending on previously known Common Vulnerabilities and Exposures (CVE) identifiers, recovering all three known CVE-fix
campaigns and surfacing two additional candidate bug families. Grounding our detectors in real fix evidence, we build four independent localizers: missing speculation barriers in C (CVE-2017-5753, Spectre~v1), missing bounds checks before array writes in C (decompression library
CVE), missing Return Stack Buffer (RSB)-stuffing macro calls in x86 assembly (CVE-2017-5715), and missing integer-overflow guards in Hand-Off Block (HOB) creation C code (surfaced by
the miner itself). All four achieve 100\% recall; precision ranges from 2.1--15.5\% on the C families to 100\% on the assembly and HOB families. We perform root-cause analysis for the C-family false positives mechanically, which attributes 77--90\% to
two intra-procedural causes, isolating the inter-procedural alias-analysis gap as a measured 15--20\% rather than an estimate. A held-out test confirms Spectre~v1 localization holds at 100\% recall on
unseen files; a fifth, independently built dictionary entry (CVE-2018-3630) shows the methodology extends to a new bug signature at low added cost; and a naive syntactic baseline recalls at most 14\% where our targeted detector
recalls 100\%. We frame these results within a broader research agenda for a unified hardware-to-firmware correctness lifecycle.
\end{abstract}



\maketitle

\section{Introduction}
\label{sec:intro}

Hardware security vulnerabilities discovered at post-silicon levels present a fundamental problem. The artifact that contains the bug cannot be changed. The Spectre and Meltdown disclosures made this concrete at scale, engineers
spent months writing hand-crafted microcode and firmware patches for vulnerabilities rooted in the speculative-execution microarchitecture of billions of already distributed processors~\cite{kocher2020spectre,lipp2020meltdown}.
Each patch had to be manually localized, synthesized, and validated, then manually re-applied across every affected firmware branch and product variant.

APR has not addressed this class of problem. State-of-the-art RTL repair systems (CirFix~\cite{ahmad2022cirfix},
RTL-Repair~\cite{laeufer2024rtl}, SRepair~\cite{liu2025srepair}, and dictionary-guided mutation operator based fixes~\cite{Mastora2026DictionaryHDLRepair}) focus on design time, producing a corrected design before fabrication. Nothing in this pipeline applies once the chip has taped out. Post-silicon, the firmware layer (BIOS/UEFI code that initializes the platform before the operating system loads) is the only accessible
portion for software-level workarounds. A different line of work on firmware \emph{hotpatching} (RapidPatch~\cite{he2022rapidpatch},
AutoPatch~\cite{salehi2024autopatch}, StackPatch~\cite{zhou2025dynamic}) automates deployment without rebooting, but explicitly depends on a human-written correct patch. As AutoPatch~\cite{salehi2024autopatch} states directly, all existing hotpatching approaches require developers to write the patch manually. Automatically localizing every vulnerable site and synthesizing a correct fix from a bug-class description before deployment has, to our knowledge, not been explored.

We ask whether the core APR methodology transfers from RTL to this firmware layer. The methodology we explore is dictionary-guided and commit-driven. The pipeline we built mines real historical fixes to build a dictionary of fix patterns, localizes bug sites matching those patterns, instantiates a repair template at each site, and validates it against the historical fix as ground truth. This mirrors how prior RTL repair work derived Design-Under-Test (DUT)-specific mutation
operators from production hardware bug fixes~\cite{Mastora2026DictionaryHDLRepair}; the question is whether the same three-step structure survives when the target language changes from Verilog to C and assembly, the oracle changes from a testbench simulator to a commit history, and the bug taxonomy shifts from
functional RTL errors to security-relevant firmware vulnerabilities.

To assess our hypothesis, we target the EDK~II (TianoCore) UEFI firmware  \cite{edk2repo}, which provides a well-documented history of CVE repairs across C-language System Management Mode (SMM) handlers and x86 assembly SMI entry stubs.
The contributions of this paper are:

\begin{itemize}
\item An \textbf{automated fix-template miner} that clusters EDK~II's
36,153-commit history by diff-content similarity, recovers all three major CVE-fix campaigns without supplying their names, and surfaces two further candidate families.

\item \textbf{Four independent localization and repair detectors}, each grounded in real fix evidence: missing speculation barriers in C (Spectre~v1); missing bounds checks in C (decompression); missing RSB-stuffing in x86 assembly; missing integer-overflow guards in HOB creation C code (miner-discovered).

\item A \textbf{root-cause triage} of the false-positive
population on the two low-precision C families, attributing 77--90\% to two
intra-procedural causes and isolating a measured 15--20\% alias-analysis
gap on the Spectre~v1 family, upgrading the precision limitation from an estimate to a diagnosis.

\item A \textbf{measured cost of dictionary extension}: a fifth,
independently built entry (CVE-2018-3630), compared against two naive
syntactic baselines that recall at most 14\% where our targeted rules
recall 100\%: two real sites have no syntactic guard signature at all,
unreachable by any such baseline regardless of tuning.

\item A \textbf{quantified manual-effort cost}: 13 distinct fix sites were
manually re-applied to stable branches 31 additional times, at up to
62~days of propagation lag, and a demonstration of automated propagation to
4 stable branches for one fix site.
\end{itemize}

\section{Background and Related Work}
\label{sec:related}

\paragraph{APR for hardware:}
Published hardware APR systems operate at design time and follow a specific structure: propose a candidate, simulate against a testbench oracle, accept or reject. CirFix~\cite{ahmad2022cirfix} applied genetic programming to synthesizable Verilog guided by testbench-output divergence; RTL-Repair~\cite{laeufer2024rtl} constrains candidates to be synthesizable via symbolic execution;
SRepair~\cite{liu2025srepair} replaces GP with LLM-guided mutation.
Recent work~\cite{Mastora2026DictionaryHDLRepair} established dictionary-guided mutation for
Verilog repair, showing DUT-specific mutation dictionaries derived from the design outperform generic operators. The insight this paper builds on: the \emph{methodology} (build a
dictionary of fix patterns, localize matching sites, instantiate a repair, validate) does not inherently require a simulation oracle or an HDL. Our approach is derived from the hypothesis that, on changing the oracle to a commit history and the target language to firmware C/assembly, the three-step structure should still apply.
To our knowledge, no existing RTL/HLS APR system addresses the scenario where the buggy artifact is a fabricated, deployed chip that cannot change.

\paragraph{UEFI firmware and the patching landscape:}
UEFI/EDK~II firmware runs in System Management Mode (SMM), a privileged x86 environment isolated from the OS; vulnerabilities there can give an attacker full platform control before any software protection layer
applies~\cite{Bazhaniuk2015smi}. Discovery-side tools (CHIPSEC's configuration and protection checks, static analysis on UEFI modules) surface vulnerabilities but produce reports, not patches. Deployment-side hotpatching automates \emph{applying} a patch without rebooting: HERA~\cite{niesler2021hera} redirects execution via ARM Cortex-M hardware breakpoints; RapidPatch~\cite{he2022rapidpatch} executes patches through a small on-device bytecode runtime so one patch release covers heterogeneous devices; AutoPatch~\cite{salehi2024autopatch} statically analyzes an official patch and generates an equivalent hotpatch automatically; StackPatch~\cite{zhou2025dynamic}
extends coverage via stack-frame reconstruction.
None of these produce the patch content itself (AutoPatch and StackPatch both state this limitation explicitly).
In the EDK~II history we study, a security engineer manually identifies each site, writes the fix, and re-applies it to every affected stable
branch, taking up to 62~days of propagation lag for some fixes (Section~\ref{sec:mining-results}).
Automated repair would slot in immediately upstream of hotpatching.

\paragraph{Commit history mining:}
Commit history mining originates with SZZ~\cite{rosa2021evaluating}, linking
bug-fix commits to introducing commits for defect prediction. In ongoing work on RTL bug mining, we applied this lineage to production RTL
repositories to derive mutation operators from actual bug fixes; as that work is unpublished and under review, we omit its results here. In this context, we adapt the same lineage differently: we cluster fixes by content similarity to surface recurring fix \emph{templates}, related to the recurring-fix-pattern literature~\cite{Martinez2013} but focused on security-relevant CVE remediations, where the fix for a hardware vulnerability class is usually
architecturally constrained.

\paragraph{Spectre and firmware-level mitigations:}
The Spectre/Meltdown disclosures~\cite{kocher2020spectre,lipp2020meltdown} are the most prominent prior works for this paper's problem: hundreds of hand-written patches, none fixable by a chip respin. RSB-stuffing addresses Spectre~v2 at SMM exit~\cite{Intel2018speculation}: the Return Stack Buffer may underflow to branch prediction on return from SMM, so the fix fills it with harmless entries before the resume-from-SMM (\texttt{rsm}) instruction. Bounds-check-bypass (Spectre~v1) addresses array accesses where the processor may speculatively execute an out-of-bounds access before a guard
resolves, requiring a speculation barrier (\texttt{AsmLfence()}) between check and use.
Both were applied manually, site by site. Our mining process quantifies the scale of that effort. The RSB fix touched 3 original sites and was then \emph{backported} (manually re-applied to older maintained release branches) 4 times, while the bounds-check-bypass fix touched 7 sites and was backported 18 times.

\section{Methodology}
\label{sec:methodology}

\begin{figure*}[t]
\centering
\definecolor{dataFill}{HTML}{F1EFE8}
\definecolor{dataLine}{HTML}{5F5E5A}
\definecolor{dataText}{HTML}{2C2C2A}
\definecolor{procFill}{HTML}{E1F5EE}
\definecolor{procLine}{HTML}{0F6E56}
\definecolor{procText}{HTML}{04342C}
\definecolor{condFill}{HTML}{FAEEDA}
\definecolor{condLine}{HTML}{854F0B}
\definecolor{condText}{HTML}{412402}
\resizebox{\textwidth}{!}{%
\begin{tikzpicture}[
  font=\small,
  bx/.style={draw, rounded corners=2pt, line width=0.4pt,
             minimum width=3cm, minimum height=1.4cm, align=center},
  data/.style={bx, fill=dataFill, draw=dataLine, text=dataText},
  proc/.style={bx, fill=procFill, draw=procLine, text=procText},
  cond/.style={bx, fill=condFill, draw=condLine, text=condText,
               minimum width=3.5cm},
  wide/.style={minimum width=3.5cm},
  flow/.style={-{Stealth[length=2mm]}, line width=0.4pt, draw=dataLine},
  feed/.style={-{Stealth[length=2mm]}, line width=0.4pt, draw=dataLine,
               dash pattern=on 2pt off 1.5pt},
  sub/.style={font=\footnotesize}
]

\draw[draw=dataLine, line width=0.4pt, dash pattern=on 2pt off 2pt,
      rounded corners=4pt] (0.6,1.3) rectangle (16.4,-1.0);
\node[anchor=west, sub, text=dataText] at (0.8,1.05)
  {\textbf{Dictionary construction, once per repository}};

\node[data] (commits) at (2.5,0) {\textbf{Commits}\\[1pt]{\footnotesize 36,153 total}};
\node[proc] (finger)  at (6.5,0) {\textbf{Fingerprint}\\[1pt]{\footnotesize 155 CVE fixes}};
\node[proc] (cluster) at (10.5,0) {\textbf{Cluster}\\[1pt]{\footnotesize 16 candidates}};
\node[data] (entries) at (14.5,0) {\textbf{Entries}\\[1pt]{\footnotesize 5 families}};

\draw[flow] (commits) -- (finger);
\draw[flow] (finger) -- (cluster);
\draw[flow] (cluster) -- (entries);

\node[data] (fw)   at (2.5,-3.8) {\textbf{Firmware}\\[1pt]{\footnotesize C and x86 asm}};
\node[proc] (loc)  at (6.5,-3.8) {\textbf{Localize}\\[1pt]{\footnotesize line scanner}};
\node[proc] (syn)  at (10.5,-3.8) {\textbf{Synthesize}\\[1pt]{\footnotesize bind, insert}};
\node[proc] (val)  at (14.5,-3.8) {\textbf{Validate}\\[1pt]{\footnotesize match or gcc}};

\draw[flow] (fw) -- (loc);
\draw[flow] (loc) -- (syn);
\draw[flow] (syn) -- (val);

\draw[feed] (13.5,-0.7) -- (13.5,-1.35) -- (6.5,-1.35) -- (6.5,-3.1);
\draw[feed] (14.5,-0.7) -- (14.5,-1.85) -- (10.5,-1.85) -- (10.5,-3.1);
\draw[feed] (15.5,-0.7) -- (15.5,-2.35) -- (14.5,-2.35) -- (14.5,-3.1);

\node[anchor=west, text=dataText] at (6.7,-2.8) {$\sigma$ signature};
\node[anchor=west, text=dataText] at (10.7,-2.8) {$\rho$ template};
\node[anchor=east, text=dataText] at (14.3,-2.8) {$\nu$ oracle};

\node[cond] (taint) at (6.5,-6.0) {\textbf{Taint filter}\\[1pt]{\footnotesize Spectre v1 only}};
\node[data, wide] (patched) at (14.5,-6.0) {\textbf{Patched source}};

\draw[flow] (taint) -- (loc);
\draw[flow] (val) -- (patched);

\end{tikzpicture}%
}
\caption{Dictionary-guided repair applied to firmware. Mining the repository's
commit history (top) yields one dictionary entry per bug family; each entry
supplies the bug signature $\sigma$, repair template $\rho$, and validation
oracle $\nu$ that drive the corresponding stage of the repair pipeline
(bottom). Gray denotes data and artifacts, teal denotes processing steps, and
amber denotes the one stage that runs for a single family rather than all of
them.}
\Description{A two-row block diagram. The top row, enclosed in a dashed box
labelled dictionary construction, shows four boxes connected left to right:
commits, fingerprint, cluster, and entries. Three dashed arrows descend from
the entries box to the second row, labelled signature, template, and oracle.
The second row shows four boxes connected left to right: firmware, localize,
synthesize, and validate. Below the second row, a taint filter box feeds
upward into localize, and validate feeds downward into a patched source box.}
\label{fig:workflow}
\end{figure*}
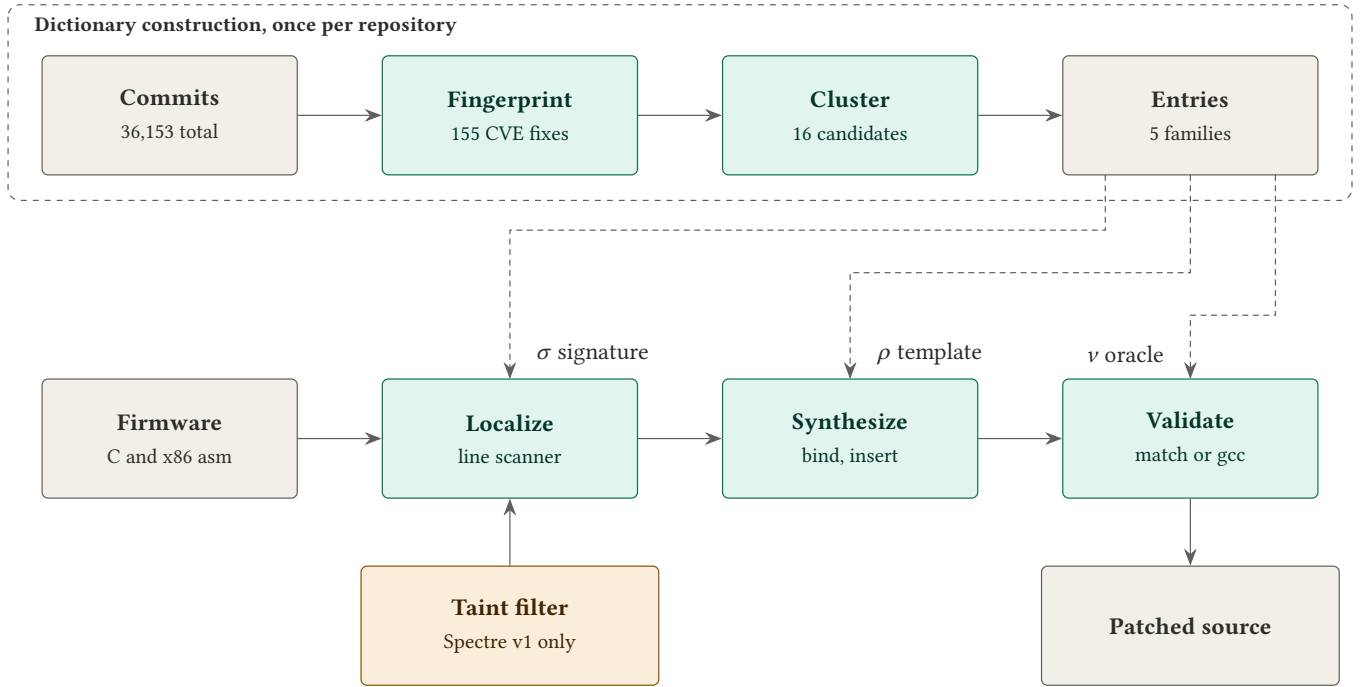

Figure~\ref{fig:workflow} summarizes the approach. Mining the commit history of
a firmware repository yields a set of dictionary entries, one per recurring bug
family. Each entry then supplies the three components that drive the repair
pipeline: a pattern that locates candidate sites, a template that instantiates
the fix, and an oracle that checks the result independently of the human fix.
The remainder of this section defines the dictionary, describes the miner that
builds it, and explains how each entry becomes a working detector.

\subsection{The Firmware Repair Dictionary}
\label{sec:dictionary}

We first make precise what we mean by a \emph{dictionary}, since the term is inherited from RTL repair and its transfer to firmware is the conceptual core of this work. In dictionary-guided RTL repair~\cite{Mastora2026DictionaryHDLRepair}, the dictionary is a DUT-specific mutation vocabulary derived from the Verilog grammar and the buggy design: it enumerates the structurally valid, category-constrained token edits available for that design, so that the repair search is confined to grammar-plausible candidates rather than a generic edit space.
We carry the same principle to firmware. A firmware repair dictionary is a set of \emph{entries}, one per bug family, where each entry is a triple $\langle \sigma, \rho, \nu \rangle$:
$\sigma$ (the \emph{bug signature}) indicates the structural pattern that identifies a vulnerable site (for example, a guard clause with no subsequent \texttt{AsmLfence} barrier); $\rho$ (the \emph{repair template}) is the code to instantiate at that site (for example, an inserted \texttt{AsmLfence ()} call); and $\nu$ (the \emph{validation oracle}) is the check that the instantiated
repair is correct independently of the human fix (for example, a structural match or a \texttt{gcc -fsyntax-only} compile).

Figure~\ref{fig:hobentry} makes this concrete for the HOB integer-overflow family. The bug signature is an alignment expression that rounds a length up to the next multiple of eight whose immediately preceding non-blank, non-comment line is not an overflow guard, so the addition can wrap the sixteen-bit length and yield an allocation far smaller than intended. The repair template is a guard on the same variable, bounded against the maximum representable value less the rounding slack, inserted before the expression. The validation oracle is the pair of checks described in Section~\ref{sec:loc-results}: a structural test that a guard on that same variable appears in the ten lines preceding the expression, and a syntax-only compile of the enclosing function against minimal type stubs.

\begin{figure}[t]
\begin{lstlisting}[style=cfix]
/* Vulnerable site matched by the bug signature */
  HandOffHob = GetHobList ();

  HobLength = (UINT16)((HobLength + 0x7) & (~0x7));
\end{lstlisting}
\hrule
\begin{lstlisting}[style=cfix]
/* Synthesized output */
  HandOffHob = GetHobList ();

  //
  // Check HobLength to avoid integer overflow in alignment.
  //
  if (HobLength > MAX_UINT16 - 0x7) {
    return NULL;
  }

  HobLength = (UINT16)((HobLength + 0x7) & (~0x7));
\end{lstlisting}
\caption{A dictionary entry applied to the HOB integer-overflow family. The
signature matches an alignment expression whose immediately preceding
non-blank, non-comment line is not an overflow guard. The template is bound to
the variable name at the located site and indented to match the surrounding
code, so the same entry applies unchanged to sites that align a differently
named length.}
\Description{Two C code fragments shown one above the other. The first shows
an assignment that rounds a length variable up to a multiple of eight, with no
guard preceding it. The second shows the same code after repair, with a
three-line comment block and a conditional guard inserted immediately before
the rounding assignment, returning early when the length is too large.}
\label{fig:hobentry}
\end{figure}

\subsection{Automated Fix-Template Mining}
\label{sec:mining}
To avoid hand-picking known CVE campaigns, we mine EDK~II's full commit history for recurring fix patterns without any prior knowledge of bug names or identifiers. The miner first collects every commit whose message references a CVE, giving 155 candidates.
For each commit, it extracts a content fingerprint made up of the lines added and removed across C, header, and assembly files, stripping unrelated metadata like copyright notices.
It then groups commits by similarity in two passes.
The first pass does an exact match on the fingerprint.
The second applies a structural-skeleton match, replacing all variable names and numeric literals with placeholder tokens, so two commits that follow the same fix shape but use different variable names still group together.
For example, one HOB fix inserts \texttt{if (HobLength > MAX\_UINT16 - 0x7)} and another inserts \texttt{if (DataLength > MAX\_UINT16 - 0x7)}.
After tokenization both become \texttt{if ( ID > ID - NUM )} and cluster together.
Any group of two or more commits is then reported as a candidate bug family.
Of the 155 candidates, 23 produced empty fingerprints, meaning their diffs contained no changes to C, header, or assembly files, and were excluded from clustering.
The remaining 132 commits yielded 16 clusters, and all three known CVE fix campaigns were recovered automatically without the miner being told their names.
The code changes themselves were distinctive enough to form the clusters.
The miner also surfaced two previously unlabeled candidates: an integer-overflow check in the Hand-Off Block creation code, and a pointer alignment applied in only one branch of a conditional.
We build full detectors for both (Sections~\ref{sec:loc-results}, \ref{sec:extensibility}).
One cluster was discarded after manual inspection revealed it grouped two commits that fixed completely unrelated bugs by coincidental structural similarity.
This reflects an inherent limitation of content-similarity clustering, which produces candidates rather than verified families.
In this work, final confirmation was manual.
Automating this confirmation step, so that spurious clusters like this are rejected without manual review, remains an open problem.

\subsection{Detector Construction}
\label{sec:synthesis}

Each dictionary entry becomes a detector through three components: a localizer that finds candidate sites, a synthesizer that instantiates the repair, and, where a family needs it, a precision filter that prunes spurious candidates. We describe each in turn. Table~\ref{tab:dictionary} states the five entries in the form introduced in Section~\ref{sec:dictionary}; the fifth is built in Section~\ref{sec:extensibility} to measure what extending the dictionary costs.

\begin{table*}[t]
\caption{The five dictionary entries. Each is a triple of a bug signature, a
repair template, and a validation oracle. Four entries share a line-local
signature; the alignment entry requires comparing two branches of a
conditional, which is why its detector is structurally different.}
\label{tab:dictionary}
\small
\setlength{\tabcolsep}{4pt}
\begin{tabular}{llp{3.9cm}p{3.3cm}p{3.9cm}}
\toprule
Entry & Lang. & Bug signature $\sigma$ & Repair template $\rho$ & Validation oracle $\nu$ \\
\midrule
Spectre v1      & C   & Guard clause with no following barrier & Insert barrier call & Structural match to fix commit \\
Decompression   & C   & Array write with no preceding bounds check & Insert bounds check & Structural match to fix commit \\
RSB stuffing    & asm & Resume-from-SMM not preceded by stuffing macro & Insert macro call & Ordinal match to fix commit \\
HOB overflow    & C   & Alignment expression with no bound on the variable & Insert overflow guard & Structural test and compile \\
Alignment       & C   & Pointer aligned in one branch only & Hoist call after the block & Structural test, compile at one site \\
\bottomrule
\end{tabular}
\end{table*}

\subsubsection{Localization}
\label{sec:localization}

We build the bug signature $\sigma$ as a lightweight line and brace-depth scanner rather than a full parser. The scanner reads each source file, tracks block structure through brace depth, and looks for the structural pattern that identifies a vulnerable site. When it finds a match, it checks whether the corresponding repair template $\rho$ is already present at that location, which happens when a duplicate copy of the code was patched in an earlier commit. If the repair is absent, it flags the site as a candidate. What changes across families is the specific pattern the scanner looks for; the three-step structure of locate, check, flag stays the same. We chose scanners over full parsers because EDK~II's headers require heavy stub-work for off-the-shelf parsers, and the fix patterns are syntactically regular enough that structured scanning suffices.

\subsubsection{Synthesis}
\label{sec:synth}

At each flagged site, the synthesizer instantiates the repair template $\rho$ directly rather than searching a candidate space. It binds the variables in the template to the names at the located site and inserts the result. Because the template is fixed, correctness reduces to whether the located site and the instantiated template match the historical fix, which we evaluate in Section~\ref{sec:eval}.

Ground truth for this comparison is derived directly from the fix commits by script, with no human labeling. For C families, it is the first real statement following each inserted guard in the fixed file. For assembly, it is the ordinal position of each resume-from-SMM (\texttt{rsm}) instruction, verified by comparing per-file counts between the buggy and fixed versions.

\subsubsection{The Precision Filter}
\label{sec:precision-filter}

The RSB stuffing, HOB, and decompression families localize cleanly, with few or no spurious candidates. The Spectre~v1 family required additional engineering. A Spectre~v1 gadget is a guard clause that checks a value before it is used in a speculative memory access, and the fix inserts an \texttt{AsmLfence()} barrier between the check and the use. The problem is that C code contains many guard clauses that match this structural shape but protect no memory access at all, so the base detector flags far more candidates than are real.

To prune these, we implement a forward taint-propagation post-filter. Starting from the identifiers in the guard condition, the filter scans forward 60 lines (set above the maximum check-to-use distance observed across fix sites, with recall insensitive to the exact window), propagating taint through alias assignments and \texttt{CopyMem} calls, and retains a candidate only if a tainted identifier eventually reaches a memory operation, such as a subscript or a pointer dereference. The intuition is that a real Spectre gadget must eventually access memory speculatively; a guard whose checked value reaches no memory operation is almost certainly not a real gadget.

\section{Empirical Evaluation}
\label{sec:eval}

We evaluate each detector on the EDK~II clone described above, organized as three questions:
\begin{itemize}
    \item How large is the manual effort our pipeline targets? (Section~\ref{sec:mining-results})
    \item How well does each detector localize and repair its family? (Sections~\ref{sec:loc-results}--\ref{sec:extensibility})
    \item Does the approach earn its complexity over naive alternatives? (Section~\ref{sec:baseline})
\end{itemize}

Ground truth throughout is extracted by script from real commit diffs, with no human labeling. We build each detector by inspecting a set of real fix commits, which we call the \emph{training set}, and evaluate generalization on \emph{held-out} files never examined during construction. No machine learning is involved; the train/held-out split tests whether a hand-written rule overfits the examples it was built from, in the same sense as a software-engineering train/test split.

\subsection{Commit Mining and Backport Cost}
\label{sec:mining-results}

Before measuring detector accuracy, we quantify the manual effort the pipeline is meant to replace. Table~\ref{tab:mining} summarizes the three CVE campaigns after deduplicating backports. We identify backports two ways: the explicit \texttt{cherry picked from commit} trailer that Git records on a re-applied commit, and, for three duplicates that carried no such trailer, a match on the commit's content fingerprint.

\begin{table*}[t]
\setlength{\tabcolsep}{4pt}
\caption{Manual remediation cost and detector accuracy across the four bug families.}
\label{tab:main}
\begin{subtable}[t]{0.42\linewidth}
\centering
\caption{EDK~II CVE remediation campaigns: distinct original fix sites
vs.\ manually re-applied backports.}
\label{tab:mining}
\footnotesize
\begin{tabular}{lrrr}
\toprule
Campaign & Sites & Bkpt & Avg \\
\midrule
Bounds check (5753) & 7 & 18 & 2.57 \\
RSB stuffing (5715) & 3 &  4 & 1.33 \\
Decompression       & 3 &  9 & 3.00 \\
\midrule
\textbf{Total} & \textbf{13} & \textbf{31} & \textbf{2.38} \\
\bottomrule
\end{tabular}

\bigskip

\caption{Naive syntactic baselines vs.\ our detector on the Spectre~v1
ground truth (7 sites); see Section~\ref{sec:baseline}.}
\label{tab:baseline}
\begin{tabular}{lrrr}
\toprule
Method & Cand & Rec. & Prec. \\
\midrule
Any comparison \texttt{if} & 110 & 0.0\%  & 0.0\% \\
Any \texttt{if} statement   & 159 & 14.3\% & 0.6\% \\
Our detector               &  79 & 100\%  & 8.9\% \\
\bottomrule
\end{tabular}
\end{subtable}%
\hfill
\begin{subtable}[t]{0.56\linewidth}
\centering
\caption{Localization results. GT: ground-truth sites; Cand: candidates
flagged; h/o (held-out) applies only to Spectre~v1 (4 unseen files, 7 GT
sites); ``unfiltered'' is the detector's three rules with no taint
post-filter, distinct from the naive baselines in
Section~\ref{sec:baseline}.}
\label{tab:eval}
\footnotesize
\setlength{\tabcolsep}{3pt}
\begin{tabular}{llrrrr}
\toprule
Bug family & Filter & GT & Cand & Rec. & Prec. \\
\midrule
\multirow{3}{*}{\shortstack[l]{Spectre\\v1 (C)}}
  & unfiltered     & 7 &  79 & 100\%  & 8.9\%  \\
  & +lex-taint     & 7 &  49 & 71.4\% & 10.2\% \\
  & +data-flow     & 7 &  72 & 100\%  & 9.7\%  \\
\midrule
Spectre v1 (h/o) & +d-flow & 7 & 329 & 100\% & 2.1\% \\
\midrule
Decompress (C)   & targeted & 9 &  58 & 100\% & 15.5\% \\
RSB (asm)        & targeted & 14 & 14 & 100\% & 100\% \\
HOB (C)          & targeted &  3 &  3 & 100\% & 100\% \\
\bottomrule
\end{tabular}
\end{subtable}
\end{table*}

The pattern that motivates automation is the gap between original sites and total effort. Across 13 distinct fix sites, engineers re-applied the same fixes to stable branches 31 additional times (avg.\ 2.38 per site), and one fix took 62~days to reach all affected branches. Every one of those 31 re-applications is a place a copy could have been missed, which is precisely the failure a detector that re-scans each branch avoids.

\subsection{Localization and Repair Results}
\label{sec:loc-results}

Table~\ref{tab:eval} reports recall and precision for all four detectors. Recall is the fraction of real fix sites the detector finds; precision is the fraction of flagged sites that are real. All four reach 100\% recall, so the discriminating axis is precision, which is where the families differ sharply.

\paragraph{Spectre v1 (C)}
With the data-flow filter, the detector finds all 7 sites at 9.7\% precision. To test whether it merely memorized the training files, we froze the detector (verifying its checksum was unchanged) and ran it on 4 files it had never seen, each 3--10$\times$ larger than the training files. It again found all 7 held-out sites, evidence that the rule generalizes rather than overfits. Precision falls to 2.1\% on these files because larger files contain proportionally more generic guards that share the vulnerable shape without being vulnerable. The lexical taint filter is shown as a negative result: it drops recall to 71.4\%, missing one site buried 10~lines deep in nested blocks and one where the checked value reaches memory through a two-hop alias chain (\texttt{CommBuffer}$\to$\texttt{CopyMem}$\to$\texttt{mVariableBufferPay\-load}$\to$\texttt{SmmVariableHeader}) unreachable by literal name matching. The forward data-flow filter recovers both.

\paragraph{Decompression (C)}
The array-write bounds-check detector finds all 9 sites across 3 files at 15.5\% precision, on its first iteration with no tuning. As a check that families genuinely need separate detectors, we ran the frozen Spectre~v1 detector on these files: it finds 0 of 9 (102 candidates, none real), confirming the two families do not share a signature.

\paragraph{RSB stuffing (x86 assembly)}
The detector flags every resume-from-SMM (\texttt{rsm}) instruction not preceded by the \texttt{StuffRsb} macro, and reaches 100\% recall and 100\% precision (14/14, no false positives) across 12 files in both MASM and NASM assembly dialects. Precision is perfect here for a structural reason, not luck: \texttt{rsm} appears only at SMM exit and nowhere else, so an unguarded \texttt{rsm} is unambiguously a real site, unlike a C comparison guard that can appear in countless benign contexts.

\paragraph{HOB integer overflow (C, miner-discovered)}
This family was surfaced by the miner rather than chosen in advance, so it is a test of the full pipeline end to end. The detector reaches 100\% recall and 100\% precision (3/3). We validate it two independent ways that do not simply re-check against the human fix: a structural oracle confirming the alignment expression is preceded by a guard on the same variable, bounded against \texttt{MAX\_UINT16} (3/3 pass on patched files, 0/3 on the unpatched originals), and a \texttt{gcc -fsyntax-only} compile of the extracted \texttt{CreateHob()} function with minimal stubs, confirming the instantiated patch is syntactically valid C (3/3 pass).

\subsection{False-Positive Root-Cause Triage}
\label{sec:triage}

The C-family precision numbers are low, so we ask a sharper question than ``how many false positives,'' namely ``why does each one occur.'' We mechanically classify every false positive from the two low-precision C families into categories. This classification runs over all candidates with no filtering, so unlike the precision filter in Section~\ref{sec:precision-filter}, it carries no risk to recall; it is a diagnostic, not a filter.

For Spectre~v1 the categories are: \emph{no-memory-use}, where the checked variable never reaches a subscript or memory-op argument in the forward window; \emph{status/guard}, where every identifier is a status, error, or NULL-style check rather than a data access; \emph{alias-gap}, where the value does reach a memory access but the link back to the SMM communication buffer crosses out of the function and cannot be traced within it; and \emph{other}. For decompression the categories distinguish loops with constant or runtime bounds, literal indices, read-only accesses, and other.

\begin{table*}[t]
\centering
\caption{False-positive root-cause triage (\% of each dataset's FPs, counted over the filtered candidate set).}
\label{tab:triage}
\small
\begin{tabular}{lrrrr}
\toprule
Dataset (FPs) & No-mem-use & Status/guard & Alias-gap & Other \\
\midrule
Spectre v1 train (59)   & 78.0\% & 3.4\%  & 15.3\% & 3.4\% \\
Spectre v1 held-out (322) & 64.3\% & 12.4\% & 19.9\% & 3.4\% \\
\bottomrule
\end{tabular}

\medskip

\begin{tabular}{lrrrrr}
\toprule
Dataset (FPs) & Const-loop & Runtime-loop & Literal & Read & Other \\
\midrule
Decompression (49) & 61.2\% & 10.2\% & 10.2\% & 8.2\% & 10.2\% \\
\bottomrule
\end{tabular}
\end{table*}

Table~\ref{tab:triage} gives the breakdown. For Spectre~v1, the two intra-procedural categories (no-memory-use and status/guard) account for 81.4\% of false positives on the training set and 76.7\% held-out, which leaves the genuinely hard inter-procedural alias-gap as a measured 15--20\%, not a vague estimate. For decompression, the provably-safe and non-write categories account for 79.6\%, with no alias component at all. Across all three datasets, 77--90\% of false positives trace to intra-procedural causes that better local analysis could remove. The residual ``other'' category on held-out Spectre~v1 (11 sites, all \texttt{CommBufferPayloadSize} guards) is not detector noise: these are genuinely attacker-facing checks that the original campaign chose not to fence, so the detector's precision floor coincides with a human exploitability judgment rather than an error.

\subsection{Extensibility: A Fifth Dictionary Entry}
\label{sec:extensibility}

To measure what adding a new bug family costs, we build a fifth entry for the miner's alignment cluster (CVE-2018-3630, surfaced in Section~\ref{sec:mining}): a pointer is assigned in both branches of an \texttt{if}/\texttt{else}, but the \texttt{ALIGN\_POINTER} call is applied in only one branch, leaving the other path unaligned. The repair moves the alignment call to a single unconditional statement after the block. This signature is harder than the other four: rather than a line-local pattern, it requires reasoning about the asymmetry between two branches, so its detector uses a character-level brace matcher that resolves nested blocks rather than a line scanner. The two known real sites served as a regression check while building it.

The detector reaches 100\% recall and 100\% precision (2/2) on both real sites (\texttt{FwVolDxe/FwVol.c}, \texttt{PeiCore/FwVol.c}), and the instantiated repair reproduces the historical fix exactly at both (the only diff is an unrelated copyright-year line). We also validate the repair independently of the human fix, as we did for the HOB entry: a structural oracle confirming exactly one unconditional alignment call on the assigned variable after the block passes at both sites, and a \texttt{gcc -fsyntax-only} compile of the extracted function passes at the \texttt{FwVolDxe} site. We did not stub the \texttt{PeiCore} site for compilation because its enclosing function depends on FFS-specific macros, so that site rests on the structural oracle and the exact match alone. Once the commit evidence was in hand, building this entry took minutes, against the days-to-weeks timescale of the manual campaigns in Section~\ref{sec:mining-results}. This is preliminary evidence that the cost of extending the dictionary scales with a bug signature's structural complexity, not with the size of the codebase.

\subsection{Naive Baseline Comparison}
\label{sec:baseline}

Finally, we check that the targeted rules earn their complexity. A natural objection to any pattern-based detector is that a much simpler pattern might work just as well, so we compare the Spectre~v1 detector against the two simplest signatures someone would reach for before building a targeted rule (Table~\ref{tab:baseline}): flagging every \texttt{if(...)} that uses a comparison operator, and flagging every \texttt{if(...)} at all. If either matched our recall, the targeted rule would not be worth its complexity.

Neither comes close. The comparison-operator baseline (110 candidates) finds zero real sites. The reason is that most real guards are not value comparisons at all but function-call checks such as \texttt{if (!EFI\_ERROR (Status))}, which test whether a prior operation succeeded before running the protected code. Our detector had to be extended to recognize this function-call form of guard, added after inspecting a real site it first missed, and it is invisible to a comparison-operator match. The any-\texttt{if} baseline (159 candidates) is broader but almost useless, finding only 1 of 7 sites at 0.6\% precision, against our detector's 7 of 7 at 8.9\%.

The sharpest evidence is structural rather than numerical. Two real sites in \texttt{SmmLockBox.c} have no \texttt{if} guard anywhere near the vulnerable line: the tainted data flows straight into a function-call parameter (\texttt{SaveLockBox(\&TempLockBoxParameterSave.Guid, ...)}). No \texttt{if}-based baseline, however carefully tuned, can ever reach these two sites, because there is no \texttt{if} there to match. Our detector reaches them because its dictionary-guided rule recognizes the known validation-function call itself, not the surrounding syntax, which is direct evidence that the rule does real work beyond generic scanning.

\subsection{Repair-Template Instantiation}
\label{sec:instantiation}

Because the repair instantiates a fixed template rather than searching a space of candidates, its correctness reduces to a single question: does the located site plus the instantiated template match what the human actually wrote. Per true-positive site, we extract the inserted block from our patched file and from the human-fixed file and compare them structurally. On training sites the match is exact, 7/7, including indentation. On held-out sites the insertion location is correct in all 7, but the inserted call is wrong in all 7: we insert \texttt{AsmLfence ()}, whereas the human wrote \texttt{MemoryLoadFence ()}, an abstraction the fix commit itself introduces (it resolves to \texttt{AsmLfence ()} in SMM and to nothing in DXE). Template instantiation reproduces a fix that has a single fixed form, but it cannot invent a new abstraction that did not exist before the fix. Localization stays correct in both cases, so the limitation is in synthesis, not in finding the site.

\subsection{Patch Propagation Experiment}
\label{sec:propagation}

The last experiment targets the backport cost from Section~\ref{sec:mining-results} directly: can the repair be propagated to stable branches automatically instead of cherry-picked by hand. For \texttt{FaultTolerantWriteSmm.c} and 4 stable branch targets (each diverged 47--49 lines from the source), we test three strategies: Strategy A applies the patch directly with \texttt{patch(1)}, B allows fuzzy context matching (fuzz=3), and C re-runs the detector independently on each branch instead of transplanting the patch. All three succeed on all 4 branches. Strategy C is the most robust, since it re-localizes on each branch and needs no matching context lines at all. This replaces 4 of the 31 manual cherry-picks for this one site. The 62-day lag from Section~\ref{sec:mining-results} belongs to a different site (\texttt{Variable.c}) that uses the \texttt{MemoryLoadFence} abstraction our pipeline does not yet synthesize, so this experiment is scoped to the correct-instantiation cases.

\section{Limitations}
\label{sec:limitations}

The limitations below bound what the present results support rather than what the method could eventually do. We are addressing several of them in ongoing work, and Section~\ref{sec:agenda} states the concrete next steps.

\begin{itemize}
\item \textbf{Validation is structural rather than semantic.} The oracles confirm that a repair is placed at the correct site, that it constrains the correct variable, and, for two entries, that the enclosing function still compiles. Establishing that the inserted code mitigates the vulnerability requires an exploit differential or a regression suite, neither of which we ran.

\item \textbf{Precision reflects the detectors as built rather than as diagnosed.} The triage in Section~\ref{sec:triage} identifies intra-procedural causes behind 77--90\% of false positives, and those findings are not yet implemented as filters. The C-family precision in Table~\ref{tab:eval} is therefore a lower bound, and we have not measured how far better local analysis would move it.

\item \textbf{Baseline comparison is limited to syntactic alternatives.} The baselines in Section~\ref{sec:baseline} establish that a trivial pattern does not suffice for this problem. We did not compare against a general-purpose static analyzer or a language model prompted with the same fix commits, so we cannot say how the detectors fare against a serious alternative.

\item \textbf{Template instantiation is bounded by the vocabulary available at mining time.} On held-out sites the insertion location is correct while the inserted call is not, because the historical fix introduced a wrapper abstraction that did not exist beforehand (Section~\ref{sec:instantiation}). Synthesizing a repair that introduces a new abstraction is a distinct problem from instantiating a known one.

\item \textbf{Detection rules and ground truth derive from the same commits.} Freezing the rules before the held-out evaluation mitigates this circularity and recall survives, though the bug families were still identified by examining fixes. A family that leaves no CVE reference in a commit message would fall outside the miner's reach.

\item \textbf{The evidence base is one codebase with few sites per family.} All experiments use EDK~II, with per-family counts ranging from 2 to 14. These are feasibility results, and extending to other firmware codebases (Section~\ref{sec:agenda}) is required before they support a generalization claim.

\item \textbf{The scope is security remediation rather than general firmware defects.} All five families are CVE remediations whose fixes are architecturally constrained, which is what makes a fixed repair template viable. Whether the methodology extends to functional bugs, where the correct fix rarely has a canonical form, remains open.
\end{itemize}

\section{Open Questions}
\label{sec:agenda}

\paragraph{Precision via inter-procedural analysis}
Intra-procedural filters cannot close the alias-analysis gap (Table~\ref{tab:triage}); whether full alias analysis reduces candidates to a practically triageable count (5--10/file rather than 60--300) remains open.

\paragraph{Patch propagation at scale}
We demonstrated propagation for one site across 4 branches; scaling to all 13 sites/31 backports is open, especially whether independent re-detection remains reliable when synthesis produces the wrong function name.

\paragraph{Build-oracle validation}
Ground-truth string comparison does not guarantee behavioral correctness; compiling synthesized patches against the public TianoCore build system and running its unit tests is a concrete next step.

\paragraph{Benchmark extension}
The 18 fix sites across five families (four evaluated, one extensibility case study) are a starting benchmark; extending to other codebases (coreboot, u-boot, BMC firmware) would broaden the generalizability claim, now validated end-to-end by both the HOB and alignment results.

\section{Discussion}
\label{sec:discussion}

Our results indicate that the dictionary-guided localize-synthesize-validate structure transfers from RTL to firmware across four structurally different bug families and two languages, with a fifth entry added at low cost. We read firmware-layer repair as one component of a broader hardware-to-firmware correctness lifecycle, in which pre-silicon RTL and HLS repair corrects a design before fabrication and firmware repair addresses the same vulnerability classes once the silicon can no longer change, with the dictionary abstraction as the shared interface across both regimes. The held-out result (100\% recall on unseen, larger files) is the property such a lifecycle would require for deployment: rules built from a handful of fix commits generalize rather than memorize.

\section{Conclusion}
\label{sec:conclusion}

We have shown a dictionary-guided APR methodology transfers from RTL
repair to firmware-layer security workarounds across four structurally
different bug families spanning C and x86 assembly, and extends to a fifth
independently built entry at low added cost.
All four evaluated families reach 100\% recall including a held-out test, our root-cause triage isolates a measured 15--20\% alias-analysis gap on the Spectre~v1 family, and template instantiation reproduces the historical fix exactly on all training sites; no single rule set covers more than one bug family, but the same methodology does.

\begin{acks}
This research was supported by the National Science Foundation (NSF) under Grant CNS-2533802
on ``Ensuring the Security of Safety-Critical Embedded Systems.''
\balance
\end{acks}

\bibliographystyle{ACM-Reference-Format}
\bibliography{references}

\end{document}